\documentclass[12pt,a4paper]{article}

\usepackage{ifthen} 
\newboolean{articletitles}
\setboolean{articletitles}{true} 
\newboolean{uprightparticles}
\setboolean{uprightparticles}{false} 
\newboolean{inbibliography}
\setboolean{inbibliography}{false} 

\usepackage{microtype}
\usepackage{lineno}  
\usepackage{xspace} 

\usepackage{graphicx}  
\usepackage{color}
\usepackage{colortbl}
\graphicspath{{./figs/}} 

\usepackage{amsmath} 
\usepackage{amssymb}
\usepackage{amsfonts}
\usepackage{mathrsfs,bm,url,color}
\usepackage{upgreek} 
\usepackage{blindtext} 

\newcommand*\patchAmsMathEnvironmentForLineno[1]{%
\expandafter\let\csname old#1\expandafter\endcsname\csname #1\endcsname
\expandafter\let\csname oldend#1\expandafter\endcsname\csname
end#1\endcsname
 \renewenvironment{#1}%
   {\linenomath\csname old#1\endcsname}%
   {\csname oldend#1\endcsname\endlinenomath}%
}
\newcommand*\patchBothAmsMathEnvironmentsForLineno[1]{%
  \patchAmsMathEnvironmentForLineno{#1}%
  \patchAmsMathEnvironmentForLineno{#1*}%
}
\AtBeginDocument{%
\patchBothAmsMathEnvironmentsForLineno{equation}%
\patchBothAmsMathEnvironmentsForLineno{align}%
\patchBothAmsMathEnvironmentsForLineno{flalign}%
\patchBothAmsMathEnvironmentsForLineno{alignat}%
\patchBothAmsMathEnvironmentsForLineno{gather}%
\patchBothAmsMathEnvironmentsForLineno{multline}%
}

\usepackage{hyperref}    
\usepackage[all]{hypcap} 

\def\ux85 {\mbox{UX85}\xspace}

\ifthenelse{\boolean{uprightparticles}}%
{

 \def\PDelta      {\ensuremath{\Delta}\xspace}                 
 \def\PXi      {\ensuremath{\Xi}\xspace}                 
 \def\PLambda      {\ensuremath{\Lambda}\xspace}                 
 \def\PSigma      {\ensuremath{\Sigma}\xspace}                 
 \def\POmega      {\ensuremath{\Omega}\xspace}                 
 \def\PUpsilon      {\ensuremath{\Upsilon}\xspace}                 
 
 \def\PB      {\ensuremath{\mathrm{B}}\xspace}                 
                  
 \def\PD      {\ensuremath{\mathrm{D}}\xspace}

 \def\PK      {\ensuremath{\mathrm{K}}\xspace}

 \def\Pb      {\ensuremath{\mathrm{b}}\xspace}                 
 \def\Pc      {\ensuremath{\mathrm{c}}\xspace}

 \def\Pi      {\ensuremath{\mathrm{i}}\xspace}

}
{

 \mathchardef\PDelta="7101
 \mathchardef\PXi="7104
 \mathchardef\PLambda="7103
 \mathchardef\PSigma="7106
 \mathchardef\POmega="710A
 \mathchardef\PUpsilon="7107
                  
 \def\PB      {\ensuremath{B}\xspace}                 
                  
 \def\PD      {\ensuremath{D}\xspace}

 \def\PK      {\ensuremath{K}\xspace}

 \def\Pb      {\ensuremath{b}\xspace}                 
 \def\Pc      {\ensuremath{c}\xspace}

 \def\Pi      {\ensuremath{i}\xspace}

}

\def\cquark    {\ensuremath{\Pc}\xspace}

\def\bquark    {\ensuremath{\Pb}\xspace}

\def\kaon  {\ensuremath{\PK}\xspace}
  \def\Kbar  {\kern 0.2em\overline{\kern -0.2em \PK}{}\xspace}

\def\Kz    {\ensuremath{\kaon^0}\xspace}
\def\Kzb   {\ensuremath{\Kbar^0}\xspace}
\def\KzKzb {\ensuremath{\Kz \kern -0.16em \Kzb}\xspace}
\def\Kp    {\ensuremath{\kaon^+}\xspace}
\def\Km    {\ensuremath{\kaon^-}\xspace}

\def\KpKm  {\ensuremath{\Kp \kern -0.16em \Km}\xspace}

\def\Dbar    {\kern 0.2em\overline{\kern -0.2em \PD}{}\xspace}
\def\D       {\ensuremath{\PD}\xspace}

\def\Dz      {\ensuremath{\D^0}\xspace}
\def\Dzb     {\ensuremath{\Dbar^0}\xspace}
\def\DzDzb   {\ensuremath{\Dz {\kern -0.16em \Dzb}}\xspace}
\def\Dp      {\ensuremath{\D^+}\xspace}
\def\Dm      {\ensuremath{\D^-}\xspace}

\def\DpDm    {\ensuremath{\Dp {\kern -0.16em \Dm}}\xspace}

\def\Bbar    {\ensuremath{\kern 0.18em\overline{\kern -0.18em \PB}{}}\xspace}

  \def\Y#1S{\ensuremath{\PUpsilon{(#1S)}}\xspace}

\def\Lbar {\ensuremath{\kern 0.1em\overline{\kern -0.1em\PLambda}}\xspace}

\def\to                 {\ensuremath{\rightarrow}\xspace}

\def\AT#1     {\ensuremath{A_{\mathrm{T}}^{#1}}\xspace}           

\def\C#1      {\ensuremath{\mathcal{C}_{#1}}\xspace}                       
\def\Cp#1     {\ensuremath{\mathcal{C}_{#1}^{'}}\xspace}                    
\def\Ceff#1   {\ensuremath{\mathcal{C}_{#1}^{\mathrm{(eff)}}}\xspace}        
\def\Cpeff#1  {\ensuremath{\mathcal{C}_{#1}^{'\mathrm{(eff)}}}\xspace}       
\def\Ope#1    {\ensuremath{\mathcal{O}_{#1}}\xspace}                       
\def\Opep#1   {\ensuremath{\mathcal{O}_{#1}^{'}}\xspace}                    

\newcommand{\tev}{\ensuremath{\mathrm{\,Te\kern -0.1em V}}\xspace}
\newcommand{\gev}{\ensuremath{\mathrm{\,Ge\kern -0.1em V}}\xspace}
\newcommand{\mev}{\ensuremath{\mathrm{\,Me\kern -0.1em V}}\xspace}
\newcommand{\kev}{\ensuremath{\mathrm{\,ke\kern -0.1em V}}\xspace}
\newcommand{\ev}{\ensuremath{\mathrm{\,e\kern -0.1em V}}\xspace}
\newcommand{\gevc}{\ensuremath{{\mathrm{\,Ge\kern -0.1em V\!/}c}}\xspace}
\newcommand{\mevc}{\ensuremath{{\mathrm{\,Me\kern -0.1em V\!/}c}}\xspace}
\newcommand{\gevcc}{\ensuremath{{\mathrm{\,Ge\kern -0.1em V\!/}c^2}}\xspace}
\newcommand{\gevgevcccc}{\ensuremath{{\mathrm{\,Ge\kern -0.1em V^2\!/}c^4}}\xspace}
\newcommand{\mevcc}{\ensuremath{{\mathrm{\,Me\kern -0.1em V\!/}c^2}}\xspace}

\def\gsim{{~\raise.15em\hbox{$>$}\kern-.85em
          \lower.35em\hbox{$\sim$}~}\xspace}
\def\lsim{{~\raise.15em\hbox{$<$}\kern-.85em
          \lower.35em\hbox{$\sim$}~}\xspace}

\def\tell1  {TELL1\xspace}
\def\ukl1   {UKL1\xspace}

\usepackage{cite} 
\usepackage{mciteplus}
\usepackage{overpic}
\usepackage{subfigure}

\usepackage[noabbrev,capitalise]{cleveref}

\usepackage{floatrow}

\newfloatcommand{capbtabbox}{table}[][\FBwidth]

\usepackage{caption}
\usepackage{longtable}
\usepackage{ifpdf}

\newcommand{\re}[2][()] {\ifthenelse{\equal{#1}{()}}{{\ensuremath{{\rm \, Re}}\left(#2\right)}}
                                                    {{\ensuremath{{\rm \, Re}}\left[#2\right]}}}
\newcommand{\im}[2][()] {\ifthenelse{\equal{#1}{()}}{{\ensuremath{{\rm \, Im}}\left(#2\right)}}
                                                    {{\ensuremath{{\rm \, Im}}\left[#2\right]}}}

\definecolor{orange}{rgb}{1,0.5,0}

\usepackage{booktabs}
\usepackage{rotating}
\usepackage{multirow}

\newcommand\snowmass{\begin{center}\rule[-0.2in]{\textwidth}{0.01in}\\\rule{\textwidth}{0.01in}\\
\vskip 0.1in Submitted to the Proceedings of the US Community Study\\ 
on the Future of Particle Physics (Snowmass 2021)\\ 
\rule{\textwidth}{0.01in}\\\rule[+0.2in]{\textwidth}{0.01in} \end{center}}

\begin{document}
\renewcommand{\thefootnote}{\fnsymbol{footnote}}
\setcounter{footnote}{1}
\begin{titlepage}

\snowmass
\vspace*{1.5cm}

{\bf\boldmath\huge
\begin{center}
High precision in CKM unitarity tests in \bquark and \cquark decays
\end{center}
}

\vspace*{0.5cm}

\begin{center}
A.~Lenz$^{1}$,
S.~Monteil$^{2}$
\bigskip\\
{\it\footnotesize
$^1$Siegen University, Theoretische Physik 1, Center for Particle Physics Siegen (CPPS), \\Walter-Flex-Str. 3, 57072 Siegen, Germany
\\
$^2$Universit\'e Clermont Auvergne, CNRS/IN2P3, LPC, Clermont-Ferrand, France 
}
\end{center}

\vspace{\fill}

\begin{abstract}
\noindent
This article surveys the important questions attached to unitarity tests of the Cabibbo-Kobayashi-Maskawa matrix, with a focus on the direct determinations of the magnitude of CKM matrix elements. The current CKM anomalies are discussed and the clear-cut prospects at LHCb and Belle II envisaged. The anticipated precision on $C\!P$-conserving and $C\!P$-violating observables at the projected upgrades of the LHCb and Belle~II experiments is examined from the point of view of the search for New Physics in $B$-meson mixing, addressing the bottlenecks in precision for the interpretation of the measurements for future experiments and highlighting the related theoretical and experimental challenges. The vibrant prospects at future $e^+e^-$ colliders running at the $Z^0$ pole and $W^+W^-$ are eventually discussed.    
\end{abstract}

\vspace{\fill}

\end{titlepage}

\pagestyle{empty}  


\renewcommand{\thefootnote}{\arabic{footnote}}
\setcounter{footnote}{0}
\tableofcontents
\cleardoublepage
\pagestyle{plain} 
\setcounter{page}{1}
\pagenumbering{arabic}


\graphicspath{{figs/}}
\allowdisplaybreaks


\section{Motivation and State of the Art} 
\label{section:physics}

\subsection{Flavour Physics in / for the SM} 
Measurements in flavour physics such as decays and mixing amplitudes were instrumental in the elaboration of the Standard Model (SM). The smallness of the branching fraction of the $K^0_{\rm L}$ dimuonic decay triggered the GIM mechanism and the prediction of the charm quark~\cite{GIM},  the existence of a third generation~\cite{Kobayashi:1973fv} was suggested from the first observation of $C\!P$ violation, measurements of $B^0-\overline{B}{}^0$ oscillations hinted at a very massive top quark~\cite{ALBRECHT1987245}, to cite few of the building blocks bringing the SM to some completeness. Flavour observables are sensitive to amplitudes (modulus and phases) involving heavy virtual particles, in a similar and complementary way as electroweak precision observables. This  makes Flavour Physics a tool for discovery as potential new heavy states can contribute to these loops in such a way that the Beyond Standard Model (BSM) energy scale probed by the Flavour and Electroweak observables exceeds by far that achievable with direct searches.       

\subsection{Importance of CKM profile: a pillar of SM} 
\begin{figure}
    \centering
    \resizebox{0.8\textwidth}{!}{\includegraphics{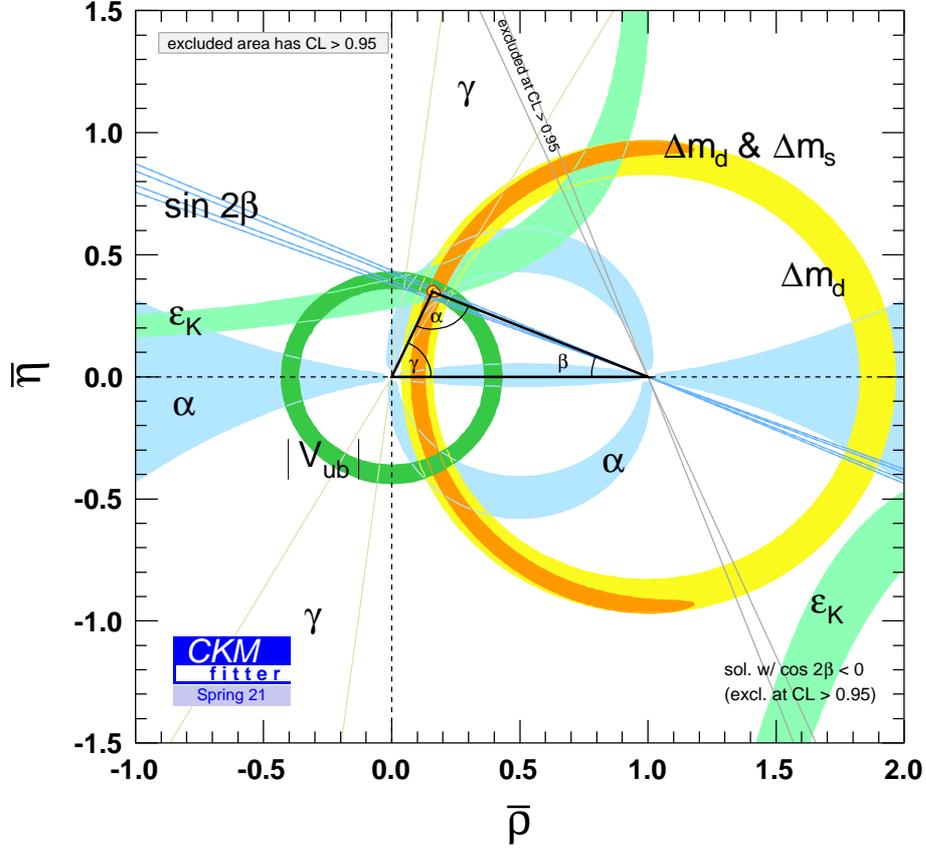}}
    \caption{Unitarity Triangle nowadays~\cite{CKMfitter2015}}
    \label{fig:ckmstateofthehart}
 \end{figure}

Though the attention is understandably attracted these days by hints of violations of lepton universality in both penguin-mediated processes and in tree decays mostly reported by the LHCb, Belle and BaBar experiments (a companion white paper is addressing this question - \cite{Guadagnoli:2022oxk}, see also  \cite{Altmannshofer:2022hfs}), the discovery power of flavour physics lies, still and as well, in $C\!P$-conserving and $C\!P$-violating observables used to constrain the CKM matrix profile and the subsequent null-tests originating from these predictions. Figure \ref{fig:ckmstateofthehart} displays the state-of-the-art of the SM CKM profile. 

The remarkable (and intriguing) consistency of flavour observables within the SM framework constitutes a pillar of the SM and even provided an indication that TeV-scale BSM physics would not be observed at the LHC. Despite this overall agreement, some anomalies within the CKM profile arise and will be discussed in this paper.               

\subsection{New Physics in mixing : a figure of merit for the experiments sensitivities}
As mentioned above, the mixing of neutral mesons has been playing a crucial role in the SM foundations and furthermore enables to access higher scales for BSM physics than the ones probed directly at LHC. The constraints on BSM Physics contributions to $B_d$ and $B_s$ mixing can then be used as a benchmark to evaluate the respective sensitivities of the future experiments. We'll follow here the methodology and results provided in a contributing white paper \cite{Charles:2020dfl}. 
This work acknowledges the prospective studies elaborated by the LHCb \cite{LHCb-Whitepaper} 
and Belle II \cite{Belle-II:2022cgf} collaborations and in the FCC-ee \cite{Bernardi:2022hny} white papers. We'll discuss in this paper more specifically the $C\!P$-conserving observables contributing to the metrology of the CKM profiles. A companion white paper will focus on the $C\!P$-violating observables.   

\begin{figure}
    \centering
    \includegraphics[width=.57\textwidth]{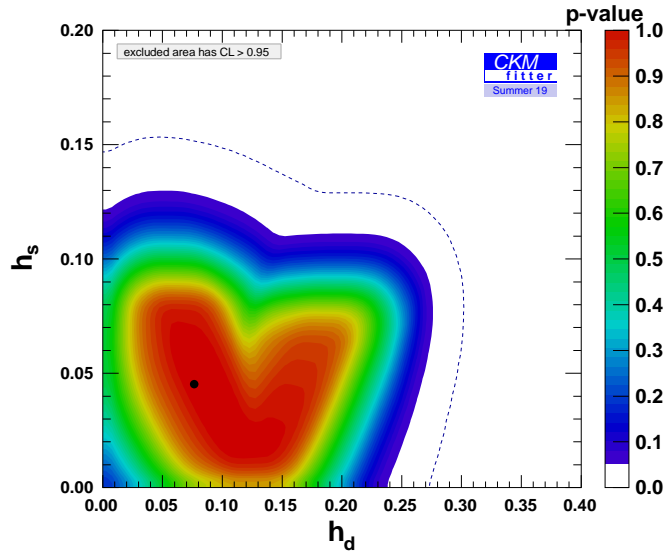}
    \caption{Exclusion curves to $h_d-h_s$ in $B_{d}$ and $B_{s}$ mixing as of \cite{Charles:2020dfl}. The SM point is at the origin. The black dot indicates the best-fit point, and the dotted curve shows the 99.7\%\,CL ($3\sigma$) exclusion contour.}
    \label{fig:hdhs_stateoftheart}
\end{figure}

Let's set the scene. Assuming the unitarity of the CKM matrix is preserved (valid in a large class of BSM models), and that the most significant BSM effects occur in observables that vanish at tree level in the SM~\cite{Soares:1992xi, Goto:1995hj, Silva:1996ih, Grossman:1997dd}
\footnote{BSM effects at tree-level are clearly not ruled out by experiment and the potential size of such effects in non-leptonic $b$-decay channels was determined in \cite{Brod:2014bfa,Lenz:2019lvd}. The allowed regions can lead to interesting effects like sizable shifts in the determination of the CKM angle $\gamma$ \cite{Brod:2014bfa} or modifications of rare $b \to s$ transitions via charm-loops \cite{Jager:2017gal,Jager:2019bgk}.}, the possible effects of heavy particles in each neutral meson system can be accounted for by two real parameters,
\begin{equation}
\label{param}
M_{12}^q = \big(M_{12}^q\big)_{\rm SM} \times
  \big(1 + h_q \, e^{2 i \sigma_q}\big)\,,
\end{equation}
where $M_{12}^q$ encodes the time evolution of the two-state neutral meson system. 
The determination of new physics (NP) contributions to meson mixing requires a precise knowledge  of the CKM elements. This is achieved by constraining the unitarity triangle (UT) apex with BSM contribution-free $\gamma$,  $|V_{ub}|$ and $|V_{cb}|$ measurements, under the adopted hypotheses above. 
The parameterisation in Eq.~(\ref{param}) is particularly convenient since the bounds on the magnitude and the phase of the BSM contributions are straightforward read off from a fit encompassing the effect many of the relevant CKM Flavour observables (any BSM contribution to $M_{12}$ is additive with respect to the SM amplitude). 
Fig.~\ref{fig:hdhs_stateoftheart} displays the status of these BSM constraints from the diverse set of observables and hadronic parameters \footnote{New lattice values from HPQCD group \cite{FermilabLattice:2021cdg} and sum rules results \cite{King:2019lal,Kirk:2017juj} would yield results in even better agreement with the SM hypothesis.} selected in \cite{Charles:2020dfl}. If the SM (origin point) passes successfully the test, one sees that BSM contributions can still be as large as $20-30\%$ that of the SM. They can be unraveled in the coming years. 
It is convenient though to assume the SM and check how the bounds are improved in the future projects. Of course this approach does not exhaust the possibilities nor opportunities offered by the experiments under scrutiny but it can serve as a useful global figure of merit of their performance.   

\section{A foreseeable experimental landscape of Flavour Physics in the next decades}

{\it 2020s decade}: Several experiments with beauty quark physics in their program are in operation or planned for the 2020s decade. The LHCb collaboration has just started the operation of its first upgrade \cite{lhcbupgrade1} during the third Run of the Large Hadron Collider (LHC) machine. ATLAS and CMS will continue to contribute significantly to the LHC flavour program in some selected areas, such as the search for the decay $B^0 \to \mu^+\mu^-$ or the measurement of the $B^0_s$ weak mixing phase $\phi_s$. The Belle~II experiment is already up and running and should integrate around 50\,ab$^{-1}$ at the horizon of the late 2020s. The LHCb detector will have recorded during the same period an integrated luminosity of 50\,fb$^{-1}$. The physics cases and the experimental commitments are fully established and we won't touch them for this prospective exercise beyond the anticipated reference point they provide.    

{\it 2030s decade}: Both experiments (LHCb and Belle II) are actively discussing further steps and upgraded detectors (see ~\cite{lhcbupgrade2} for instance), to record about 6 and 5 times larger samples during the 2030s decade, respectively, shaping the field in this timeline. Most of the $C\!P$-violating observables contributing to the study of the CKM profiles are theoretically pristine and will greatly benefit from those upgrades. For instance, the $\gamma$ angle measurement in $B^- \to DK^-$ decays  is free from theory uncertainty within the SM (smaller than $10^{-6}$ relative~\cite{Zupan:2011mn}), which ensures a continued relevance for their measurements throughout the coming decades; this will be developed in a companion white paper \cite{CPV-Overview-Whitepaper}.  This is by contrast not the case for many $C\!P$-conserving observables. There, hadronic parameters (modelling strong interaction amplitudes) do limit the precision of the predictions and their improved determination is required if one wants to maximally benefit of the increase of integrated luminosity. We will touch in this article the precision that has been anticipated by the theory community (in particular for the lattice computations) on sensitive parameters for the electroweak interpretation. 

{\it 2040s decade}: The emergence of $e^+e^-$ circular collider projects to accurately study the Brout-Englert-Higgs boson properties as well as the other relevant electroweak thresholds ($Z$, $WW$, $t \overline{t}$) opens a new appealing perspective for Flavour Physics at large and $b$-physics in particular. The statistics of $Z$ decays foreseen to be collected with the global CERN-hosted FCC-$ee$ machine project provides about 15 times more $B$ mesons than the Belle II experiment at completion. As for the LHCb experiment, all $b$-flavoured particles will be (abundantly) produced and will experience a significant boost (they are produced from a $Z$ decay), which can be decisive to measure rare processes such as $b$-hadron decays involving $\tau$ leptons in the final state \cite{Guadagnoli:2022oxk}. An $e^+e^-$ circular collider project running at the $Z$ pole (but also at and above $WW$ threshold) gathers many of the virtues of LHCb and Belle II experimental environments and therefore provides a natural perspective for the Flavour program. A similar project (CEPC) is studied in China. Since the Flavour program was embodied in the FCC project from the origin and that the luminosity target at the $Z$ pole is superior yet, we'll consider the FCC projections in this paper. 

\section{Direct determination of CKM matrix elements and discussion of CKM anomalies}
\subsection{$V_{cb}$}
Semileptonic $b$-hadron decays enable in principle a clean and direct determination 
of   $V_{cb}$ and $V_{ub}$. Using data for inclusive semileptonic $B \to X_c l \nu$ decays
from BaBar and Belle \cite{BaBar:2014omp} one can extract $V_{cb}$
\cite{ParticleDataGroup:2020ssz}
using the framework of the heavy quark expansion \cite{Lenz:2014jha}
as
\begin{equation}
    |V_{cb}|^{\rm incl., PDG} = (42.2 \pm 0.8) \cdot 10^{-3} \, .
\end{equation}
Including the recently calculated NNNLO-QCD corrections \cite{Fael:2020tow}to the free quark decay and NLO-QCD corrections to the Darwin term \cite{Mannel:2019qel}
one finds \cite{Bordone:2021oof} even smaller uncertainties
\begin{equation}
    |V_{cb}|^{\rm incl., 2022} = (42.16 \pm 0.51) \cdot 10^{-3} \, .
\end{equation}
Exclusive semileptonic decays measured in the decay channels 
$B \to D \ell\nu$, $B \to D^\ast\ell\nu$ and $B_s \to D_s^\ast\ell\nu$,
at Babar, Belle, CLEO and LHCb \cite{HFLAV:2019otj}
yield in combination with 
lattice results and sum rule determinations of the arising form factors
typically smaller values for $V_{cb}$:
\begin{equation}
    |V_{cb}|^{\rm excl., PDG} = (39.5 \pm 0.9) \cdot 10^{-3} \, .
\end{equation}
In contrast to the inclusive extraction, where only one group is performing the fit,
many more theory groups are working on exlcusive decays, leading to a larger spread in the corresponding $V_{cb}$ determinations; recent
lattice results 
(FNAL/MILC \cite{FermilabLattice:2021cdg},
HPQCD \cite{Harrison:2021tol}), 
combinations \cite{Martinelli:2021onb} of lattice \cite{MILC:2015uhg}
and unitarity
and sum rule determinations
(LCSR1 \cite{Gao:2021sav},
LCSR2 \cite{Bordone:2019vic}) 
yield in combination with data from BaBar, Belle and LHCb e.g.
\begin{eqnarray}
    |V_{cb}|^{\rm excl.} 
        & = & 
    \left\{
 \begin{array}{lll}
    (38.40 \pm 0.74) \cdot 10^{-3}
    &
    {\rm FNAL/MILC, BaBar, Belle}
    & 
    B \to D^\ast\ell\nu \, ,
\\
        (40.3 \pm 0.8 ) \cdot 10^{-3} 
        & 
        {\rm LCSR2 \, and \, lattice , BaBar, Belle}
        & 
        B \to D^{(*)}\ell\nu \, ,
\\
 (40.3 \pm 1.7 ) \cdot 10^{-3}
 & 
 {\rm LCSR1, BaBar}
 & B \to D\ell\nu \, ,
\\
        (41.0\pm 1.3 ) \cdot 10^{-3} 
        & {\rm LCSR1, Belle}
        & B \to D\ell\nu \, ,
\\
    (41.0 \pm 1.2) \cdot 10^{-3}
    &
    {\rm lattice + unitarity, Belle}
    & 
    B \to D\ell\nu \, ,
\\
        (42.2 \pm 2.3) \cdot 10^{-3} 
        & 
        {\rm HPQCD, LHCb}
        &B_s \to D_s^\ast\ell\nu \, .
        \end{array} 
\right.
\end{eqnarray}
More details about the exclusive vs. inclusive puzzle, in particular a discussion of different parameterisations of the form factors used for the extraction, can be found in \cite{Gambino:2020jvv}.

The $V_{cb}$element represents a crucial input parameter for the determination of the CKM matrix assuming unitarity and many SM predictions are very sensitive to the quoted precision of  $V_{cb}$.
Within the SM one can e.g. fix the CKM elements $V_{tx}$
from $|V_{us}|$, $|V_{cb}|$, $|V_{ub}|$ and $\gamma$ 
via the exact relations: 
\begin{eqnarray}
  V_{tb} V_{ts}^* & = & - c_{12} \frac{\sqrt{1-|V_{ub}|^2-V_{cb}^2}}{\sqrt{1-|V_{ub}|^2}} V_{cb}
  - s_{12} \frac{1-|V_{ub}|^2-V_{cb}^2}{\sqrt{1-|V_{ub}|^2}} V_{ub} \, ,
  \label{Vts}
  \\
  \frac{V_{td}^*}{V_{ts}^*} & = & \frac{s_{12} V_{cb} - c_{12} \sqrt{1-|V_{ub}|^2-V_{cb}^2} V_{ub}}
                                      {-c_{12} V_{cb} - s_{12} \sqrt{1-|V_{ub}|^2-V_{cb}^2} V_{ub} }
\label{VtdoverVts}
\end{eqnarray}
with
\begin{equation}
   s_{12} = \frac{\frac{V_{us}}{V_{ud}}}{\sqrt{1 + \frac{V_{us}^2}{V_{ud}^2}}} \, ,
  \hspace{1cm}
  c_{12} = \frac{1}{\sqrt{1 + \frac{V_{us}^2}{V_{ud}^2}}} \, ,
  \hspace{1cm}
   V_{ub}  =   |V_{ub} |e^{-i \gamma} \, 
  \end{equation}
and thus predict quantities like the mass difference of neutral $B$ mesons, $\Delta M_q$ or rare penguin decays like $B_q \to \mu  \mu $. This strategy can also be turned around and the loop induced constraints can be used to constrain   $V_{cb}$, see e.g. \cite{Altmannshofer:2021uub,King:2019rvk,DiLuzio:2019jyq}, to find
\begin{eqnarray}
|V_{cb}|^{\Delta M_q} & = & (41.6 \pm 0.7) \cdot 10^{-3} \, ,
\label{Eq:VcbDeltaM}
\\
|V_{cb}|^{\rm Meson \, mixing} & = & (42.6 \pm 0.5) \cdot 10^{-3} \, ,
\label{Eq:VcbMM}
\\
|V_{cb}|^{\rm rare \,  decays} & = & (37.3 \pm 1.0) \cdot 10^{-3} \, ,
\label{Eq:Vcbrare}
\end{eqnarray}
where meson mixing clearly favours the inclusive
value, while rare decays shows a tendency for the
low exclusive results, albeit the latter extraction is not really conclusive, since low $q^2$ data prefer very small values of $V_{cb}$, large  $q^2$ data intermediate values and decays like $ b \to s \gamma$ prefer larger values. The high precision in
Eq.(\ref{Eq:VcbDeltaM}) and Eq.(\ref{Eq:VcbMM})
is triggered by some recent improvement on the non-perturbative input for meson mixing stemming from lattice 
\cite{FermilabLattice:2016ipl,Boyle:2018knm,Dowdall:2019bea}
and sum rules
\cite{Grozin:2016uqy,Kirk:2017juj,King:2019lal}, 
with averages of both methods presented in 
\cite{DiLuzio:2019jyq}. 
The current and foreseen programs at LHCb and Belle II, in conjunction with more accurate predictions for hadronic parameters shall hopefully shed light on this puzzle. In a further future at $e^+e^-$ circular colliders, one  could imagine to determine $|V_{cb}|$ thanks to the decay $B_c^+ \to \tau^+ \nu$ \cite{Amhis:2021cfy,Zheng:2020ult}. That determination requires however the knowledge of the production fraction of the $B_c$ meson at the $Z$ pole, which can be a limiting factor. A more promising avenue avenue using on-shell $W$ decays is explored in the last section of this paper.
\subsection{$V_{ub}$}
A similar problem arises in the determination of $V_{ub}$.
PDG \cite{ParticleDataGroup:2020ssz} quotes again quite different values for the inclusive (see \cite{Lange:2005yw,
Gambino:2007rp,Andersen:2005mj}) and exclusive 
(e.g.  \cite{FermilabLattice:2015mwy,  Flynn:2015mha, Dalgic:2006dt,Colquhoun:2022atw}) determination based on data from Babar and Belle \cite{HFLAV:2019otj}:
\begin{eqnarray}
    |V_{ub}|^{\rm incl., PDG} & = & (
    4.13 \pm 0.26) \cdot 10^{-3} \, ,
    \\
    |V_{ub}|^{\rm excl., 2022} & = & 
    (3.70 \pm 0.16) \cdot 10^{-3} \, .
\end{eqnarray}
 However, now the effect on quantities like $\Delta M_q$ arising from this difference is minor \cite{King:2019rvk} compared to the ambiguities arising from different values of $|V_{cb}|$, therefore will not discuss the current determinations in more detail. Moreover the inclusive determination is now disturbed by large $b \to c$ backgrounds which neccesitates the introduction of shape functions \cite{Neubert:1993ch,Bigi:1993ex}. An improvement in the precision of the inclusive determination of $V_{ub}$ might be achievable by an experimental determination of the shape functions, see e.g. Ref. \cite{Bernlochner:2020jlt}. The exclusive determination relies again on the non-perturbative calculation of form factors via lattice QCD and/or LCSR. Interestingly 
 different $b \to u$ transitions (like $B^+ \to \tau^+ \nu_\tau$, $B^0 \to \pi^- l^+ \nu$, $B \to X_u l \nu$ and $\Lambda_b \to p \mu^- \bar{\nu}$ ) are affected differently  
 by e.g. right-handed currents stemming from BSM contributions, see e.g. Ref. \cite{Albrecht:2017odf}.
 
 The ratio $|V_{ub}/V_{cb}|$ can also be determined from  $B_s \to K l \nu / B_s \to D_s l \nu$ using input from lattice
 \cite{Bouchard:2014ypa,Flynn:2015mha, FermilabLattice:2019ikx, Harrison:2021tol}
or sum rules \cite{Khodjamirian:2017fxg}. Further information on that ratio can be gained from 
 baryonic decays like $\Lambda_b \to p \mu \bar{\nu}$ and $\Lambda_b \to \Lambda_c \mu \bar{\nu}$ 
 \cite{LHCb:2015eia} combined with lattice evaluations  \cite{Detmold:2015aaa}.
 
 A determination of $|V_{ub}|$ from $B^+ \to \tau^+ \nu$ depending only on the decay constant $f_B$ yields larger uncertainties \cite{ParticleDataGroup:2020ssz}
\begin{eqnarray}
    |V_{ub}|^{\tau \nu} & = & (
    4.13 \pm 0.26) \cdot 10^{-3} \, .
\end{eqnarray}%
 As for $V_{cb}$, significant progress is expected from the LHCb \cite{LHCb-Whitepaper} and Belle~II \cite{Belle-II:2022cgf} current and future programs. Prospects at future $e^+e^-$ colliders must be quantitatively assessed but a measurement at the percent level of the $B^+ \to \tau^+ \nu$ decay  seems a plausible and promising avenue.%
\subsection{Cabibbo Anomaly}
 Due to unitarity the elements of the first row of the CKM matrix
 obey the relation
\begin{equation}
  |V_{ud}|^2 + |V_{us}|^2 + |V_{ub}|^2 = 1  \, .
\end{equation}
The CKM element $V_{ud}$ is determined precisely from super allowed nuclear beta decay, neutron decay, and pion beta decay and one obtains \cite{ParticleDataGroup:2020ssz}
(see also \cite{Seng:2021syx})
\begin{equation}
  V_{ud} =  0.97373(31) \, ,
\end{equation}
where the dominant uncertainty stems from the nuclear structure.
The CKM element $V_{us}$ can be  determined  from kaon decays, hyperon decays, and tau decays and one obtains \cite{ParticleDataGroup:2020ssz} a weighted average of  $Kl3$ and $K \mu 2$ decays of
\begin{equation}
  V_{us} =  0.2243(4)\, .
\end{equation}
 Note that the results of  $Kl3$ and $K \mu 2$  differ by 3 standard deviations, see e.g.
 \cite{Seng:2022wcw}.    
 For testing unitarity of the first row the uncertainties in $V_{ub}$ are irrelevant.
PDG finds \cite{ParticleDataGroup:2020ssz} with a conservative error estimate
\begin{equation}
  |V_{ud}|^2 + |V_{us}|^2 + |V_{ub}|^2 = 0.9985(6)(4) \, ,
\end{equation}
which indicates a two sigma deviation from unitarity, while other studies
\cite{Kirk:2020wdk,Belfatto:2019swo,Grossman:2019bzp,Aoki:2021kgd} find deviations of 
the order of four sigma. The origin of such a discrepancy could be rooted in
an underestimation of the uncertainties in nuclear decays, in the lattice
determination of the kaon form factors or in new physics. 
Continued efforts in the strengthening of the hadronic predictions are definitely desirable.
\subsection{$V_{cd}$ and $V_{cs}$}
 Corresponding tests of unitarity via the second row of the CKM matrix
\begin{equation}
  |V_{cd}|^2 + |V_{cs}|^2 + |V_{cb}|^2 = 1  \, .
\end{equation}
turn out to be inconclusive due to the larger uncertainties in the large CKM elements
$V_{cd}$ and $V_{cs}$. These elements are currently determined from semi-leptonic decays $D \to \pi l \mu$, $D \to K l \nu$ and leptonic decays
$D^+ \to \mu \nu$, $D_s^+ \to \mu \nu$ and $D_s^+ \to \tau \nu$,
to obtain \cite{ParticleDataGroup:2020ssz}
   \begin{eqnarray}
    |V_{cd}| = 0.221(4) \, ,
    &&
    |V_{cs}| = 0.987(11) \, .
       \end{eqnarray}
The extraction from semi-leptonic decays requires the knowledge of form-factors, while
the extraction from leptonic decays requires the very well-known decay constants - thus the latter decays will provide precise values with continuing experimental progress
and finally enable more stringent unitarity tests of the second row. 
Future lattice studies aiming to achieve a higher precision will have to include also
QED correction, see e.g. the discussion in Ref.\cite{Boyle:2022uba}.
A very precise determination of $V_{cs}$ using lattice QCD and incorporating QED effects was recently presented by the HPQCD collaboration \cite{Chakraborty:2021qav}
   \begin{eqnarray}
    |V_{cs}| &= & 0.9663(79) \, .
       \end{eqnarray}
 The experimental precision in the determination of $V_{cd}$ and $V_{cs}$ will be considerably increased at a  future super-tau-charm-factory, see e.g. the white paper of the Chinese project \cite{Cheng:2022tog}.
For the further future, on-shell $W$-boson decays at $e^+e^-$ circular colliders shall provide the ultimate precision on the determination of the $|V_{cs}|$ matrix element \cite{Bernardi:2022hny}.
\subsection{Direct determination of $V_{tx}$}
 Combinations of CKM elements of the last row can be determined from loop processes, 
 see e.g. \cite{Dowdall:2019bea,King:2019rvk}, while direct information of an individual
 top-CKM element is so far only available from single-top production, yielding
 \cite{ParticleDataGroup:2020ssz}
    \begin{eqnarray}
    |V_{tb}| = 1.013(30) \, .
       \end{eqnarray}
 Continued improvement on the $|V_{tx}|$ matrix element precision is expected at HL-LHC \cite{Cerri:2018ypt}. The operation of FCC-$ee$ at the top-pair production threshold provides as well desirable opportunities in this area.     


\section{Projected sensitivities for NP in mixing in the next two decades} 

Some comments are in order about the key observables at play in these global analyses and the corresponding experimental sensitivities. We discuss in this section the LHCb and Belle II prospects and will dedicate a section about the future $e^+e^-$ circular colliders related prospects.   

\begin{itemize}
    \item 
    The CKM matrix elements $|V_{ub}|$ and $|V_{cb}|$ are important ingredients to determine the UT apex. 
    They shall be  determined with a precision of $1.2\%$ and $1.4\%$, respectively, in exclusive semi-leptonic 
    decays by the Belle~II experiment with 50 1/ab \cite{Belle-II:2018jsg,Belle-II:2022cgf}. The presence of the 
    neutrino is more difficult to constrain in the LHCb spectrometer but the unique study of the semileptonic 
    decays of $B_s$ and $\Lambda_b$ particles brings additional information \cite{lhcbupgrade2,LHCb-Whitepaper}. The ratio of matrix elements $|V_{ub}|/|V_{cb}|$ is expected to be ultimately measured at $3\%$ and $1\%$ precision with 50 1/fb and 300 1/fb, respectively. 
    The possibility to collect several $10^8$ $W$ decays at FCC-$ee$ offers the opportunity to measure  $|V_{cb}|$ from on-shell $W^+ \to c \bar{b}$ decays
    (\cite{Monteil:2021ith} and references therein). 
    The total uncertainty in the latter quantity is estimated to be better than the one extracted from the future upgrade of Belle~II.
    \item In order to interpret the $B$-meson mixing measurements, non-perturbative hadronic parameters must be determined with lattice QCD computations. These are planned to be computed with a precision  better than $1\%$ \cite{Belle-II:2018jsg,Cerri:2018ypt}. Those projections will be addressed in a dedicated section. 
    \item Eventually, the $C\!P$-violating CKM angles $\alpha, \beta, \beta_s, \gamma$, described in a companion white paper \cite{CPV-Overview-Whitepaper}, shall all be measured with a subdegree uncertainty \cite{Belle-II:2018jsg,Abada:2019lih,Cerri:2018ypt,lhcbupgrade2,Bernardi:2022hny,Belle-II:2022cgf,LHCb-Whitepaper}.  
    
\end{itemize}

\begin{figure}
\centering
\minipage{0.32\textwidth}
  \includegraphics[width=\linewidth]{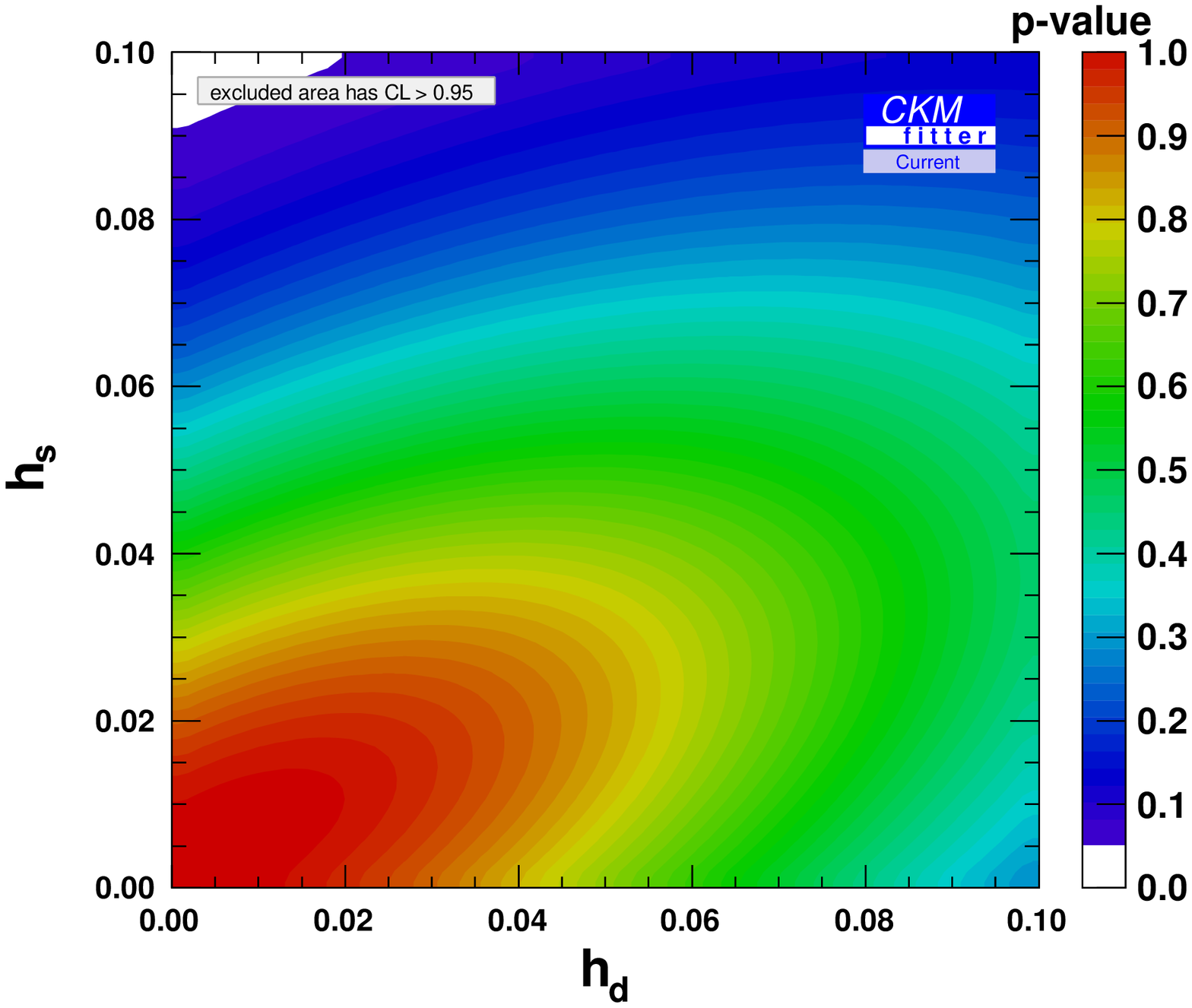}
\endminipage\hfill
\minipage{0.32\textwidth}
  \includegraphics[width=\linewidth]{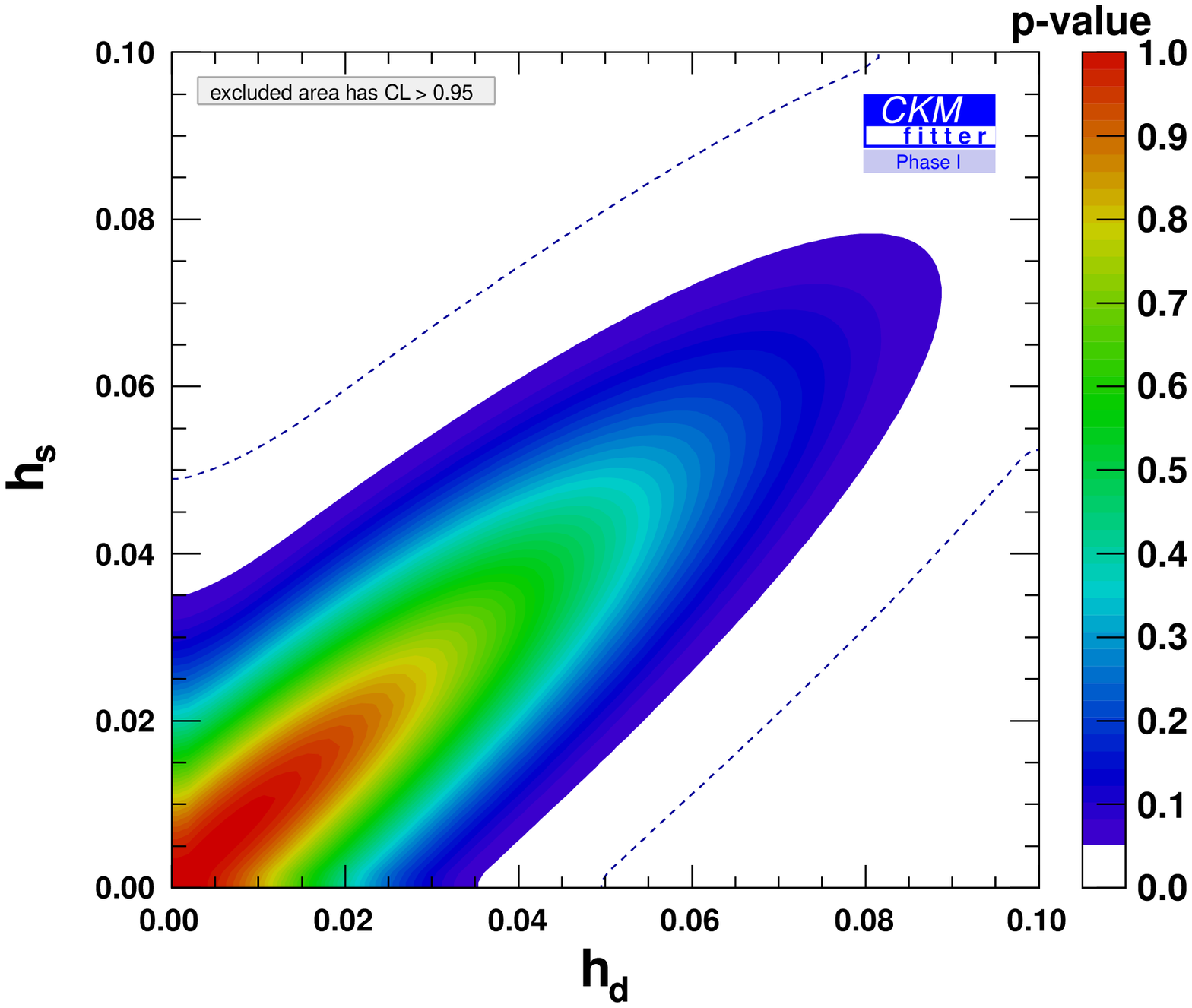}
\endminipage\hfill
\minipage{0.32\textwidth}
  \includegraphics[width=\linewidth]{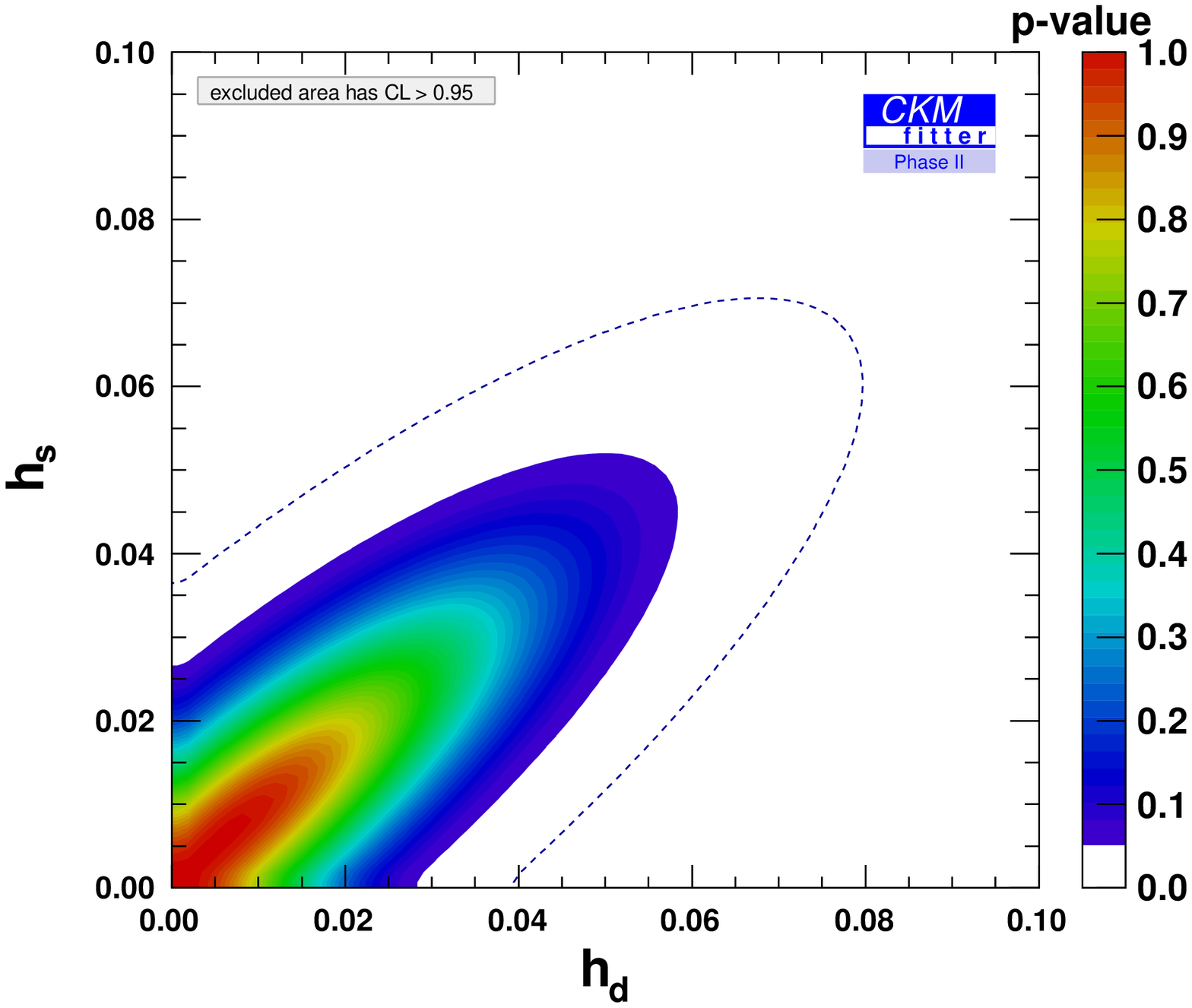}
\endminipage
\caption{
Sensitivities to the magnitude of BSM contributions in $B_{d}$ and $B_{s}$ mixing. From left to right,  one finds the current sensitivity, the anticipated constraints after LHCb Upgrade I and Belle II operations and finally those after the foreseen upgrades of Belle II and LHCb experiments. Dotted curves show the 99.7\%~C.L. ($3\sigma$) contours, assuming SM $h_d=h_s=0.$   Taken from Ref.~\cite{Charles:2020dfl}.
}
\label{fig:hdhs}
\end{figure}

%

The complete set of expected bounds on the energy scale of BSM operators, assuming consistency with the SM, can be found in \cite{Charles:2020dfl}. We report here instead in Fig.~\ref{fig:hdhs} the evolution of the constraints on the magnitudes of the potential BSM contributions at the different timescales of the future projects. 
We take note that the constraints on the magnitudes of the BSM amplitudes do improve by a factor larger than 3 w.r.t. the state of the art when considering the 2030s decade. By contrast, the bounds are not improving along with the luminosity at the horizon of 2040s decade. Two bottlenecks have been identified in \cite{Charles:2020dfl}: 
\begin{itemize} 
\item the precision anticipated for the computations of the hadronic parameters of the oscillation frequencies (namely decay constants, bag parameters, and perturbative QCD corrections) are limiting the global fit precision. As in the case of leptonic decays the inclusion of QED corrections will be mandatory to further improve the precision  \cite{Boyle:2022uba,Kronfeld:2022rjy}.   
\item maybe less expectedly, the knowledge of the matrix element $|V_{cb}|$ becomes a limiting factor when considering the LHCb and Belle II ultimate upgrades. The reason is that $|V_{cb}|$ controls the normalisation of all the sides of the UT. 
\end{itemize}

It is therefore necessary in order to maximally benefit from the anticipated precision at the future upgrades of LHCb and Belle~II, in particular for the $C\!P$-violating observables, to improve on the accuracy of the relevant lattice QCD parameters and the magnitude of the CKM matrix element $|V_{cb}|$, beyond what has been anticipating.   
In this analysis, the central values of the different observables are shifted to match the SM expectation; it is useful to provide quantitative assignments and identify the bottlenecks.  Obviously, this is a conservative approach and BSM contributions can be unraveled all the way long, alleviating the conclusions about LQCD parameters and $|V_{cb}|$.    

\section{Theory challenges }
Potential improvements in theory predictions can be obtained in several fields: 
\begin{itemize}
\item Perturbative corrections: Perturbative QCD made huge progress in recent years, see as an example the NNNLO-QCD corrections to inclusive semileptonic $b$ decays \cite{Fael:2020tow}. Future higher-order corrections will increase the precision of more observables, e.g. NNLO-QCD corrections to the mass difference of neutral mesons. Besides "simply" adding more loops, there are, however, conceptual issues, like an appropriate definition of quark masses, that require further investigations and that will reduce theory uncertainties.
\item Non-perturbative determinations of decay constants: these parameters are currently very well determined via lattice simulations, see \cite{Aoki:2021kgd}. For further reduction of theoretical uncertainties, QED corrections will have to be taken into account properly.  
Nevertheless one can study ratios like $Br(B_s \to \mu \mu)/\Delta M_s$ or  $Br(B^+ \to \tau \nu)/\Delta M_d$  where the dependence on the decay constants (and in the case of the first ratio the CKM elements dependence) cancels out  within the SM. As stated above for the study of leptonic decays and mixing a further improvement in precision requires the incorporation of QED effects, see Ref. \cite{Boyle:2022uba,Kronfeld:2022rjy}.
\item Non-perturbative determinations of form factors can be done via LCSR and lattice
simulations, which still can be systematically improved, see e.g. the lattice white paper contributions \cite{Boyle:2022uba,Kronfeld:2022rjy,Boyle:2022ncb}. In case of neutral current
transitions there are in addition sizable non-local contributions, where we need some new conceptual developments in order to drastically improve the precision. 
\item Non-perturbative determinations of mixing matrix elements were for quite some time mostly available from lattice QCD. Here the two most recent ones from FNAL/MILC 
\cite{FermilabLattice:2016ipl} (2+1) and HPQCD \cite{Dowdall:2019bea}  (2+1+1) differ quite visibly. FLAG quotes \cite{Aoki:2021kgd}
\begin{eqnarray}
f_{B_s} \sqrt{\hat{B}} & = & 274(8) \,  \rm MeV \, \, (N_f = 2+1) \, ,
\\
f_{B_s} \sqrt{\hat{B}} & = & 256.1(5.7) \,  \rm MeV \, \, (N_f = 2+1+1) \, .
\end{eqnarray}
A further convergence of  the lattice results is mandatory in order to achieve the planned future relative precision of $1 \%$ \cite{Belle-II:2018jsg,Cerri:2018ypt}; an independent lattice determination of the ratios of $B_s$ and $B_d$ mixing was presented in \cite{Boyle:2018knm}.
Sum rules provide a completely different method for determining the Bag parameter contributing to the mixing matrix elements. $B_d$ mixing was studied with HQET sum rules in \cite{Grozin:2016uqy,Grozin:2017uto,Kirk:2017juj,Grozin:2018wtg},
while $B_s$ mixing was studied in the same framework in \cite{King:2019lal}.
Averages of lattice and HQET sum rules were presented in  \cite{DiLuzio:2019jyq}. The sum rules estimates can also be further improved by determining $1/m_b$ corrections to the strict HQET limit, by determining NNLO-QCD corrections to the QCD-HQET matching or by considering the sum rule in full QCD.
\end{itemize}
A very interesting new theory development is the determination of inclusive semileptonic 
decays directly on the lattice: \cite{Hashimoto:2017wqo,Hansen:2019idp,Bulava:2019kbi,Gambino:2020crt,Bulava:2021fre,Gambino:2022dvu}, which could provide further insights in the precise value of the crucial parameter $V_{cb}$.

\section{A possible future for  Flavour Physics -- $e^+e^-$ future circular colliders.}

Flavour-physics experiments using $e^+e^- \to \Upsilon(4S) \to b\overline{b}$, such as Belle~II, benefit from experimentally clean final-state events and fully efficient trigger over a large geometrical acceptance. A $pp \to b\overline{b} X$ experiment, such as LHCb, produces all species of heavy-flavoured hadrons with a high boost. Experiments relying on $e^+e^- \to Z^0 \to b\overline{b}$ production, as would be the case at FCC-ee or CEPC, are combining these  advantages. The invincibly large $b\overline{b}$ production cross-section at LHC is partially mitigated by the high luminosity that is foreseen in particular at FCC-ee and the characteristics of the $Z^0$ environment allows for studies that are complementary to (or more sensitive than)  those foreseen at LHCb and its upgrades. Table~\ref{tab:flavouryields} gives the produced yields in $5 \times 10^{12}$ $Z^0$ decays for each $b$-hadron species, as well as those for charm and $\tau$ particles. Yields of $B^0$ and $B^+$ are therefore around fifteen times larger than those expected at the completion of Belle~II. A further discussion about the advantages of the $Z^0$ environment and a review of flavour physics opportunities including those sketched in this article are provided in \cite{Monteil:2021ith}. 

Flavour possibilities are not restricted to the $Z^0$ pole physics. Several $10^8$ $W$-boson decays can be registered at and above the $W^+W^-$ threshold at FCC-$ee$ and thanks to excellent vertexing capabilities, one could select events with high-purity $b$-tagged and $c$-tagged jets ~\cite{Behnke:2013xla}. 
It is then conceivable to improve the knowledge of $|V_{cb}|$ up to an order of magnitude with respect to the expected precision at the completion of the LHCb and Belle~II upgrades operations.  Very precise measurements of $|V_{cs}|$ are as well at reach using on-shell $W$ decays. Closing on the figure of merit defined from BSM contributions in mixing, the future $e^+e^-$ circular colliders can therefore improve significantly on the  anticipated bounds at the previous generation of machines, as soon as their luminosity at the $Z^0$ pole and $W^+W^-$ equates or exceeds the one targeted at FCC-$ee$. An assessment of the precision that could be reached at this horizon in the prediction of the hadronic parameters will concurrently be desirable. Most of the measurements will remain  statistically limited and the flavour program at large would benefit of larger samples. About a factor of two more events can be recorded if the design of the FCC-$ee$ would comprise four interaction-points layout as is currently studied. 

\begin{table}[htbb]
\caption{Yields of heavy-flavoured particles produced at FCC-ee for $5 \times 10^{12}$ $Z^0$ decays. Production fractions at the $Z^0$ pole are taken from ~\cite{HFLAV:2019otj,ParticleDataGroup:2020ssz}. Identical rates are obtained for charge conjugate states. \vspace*{0.1cm}} 
\centering
\begin{tabular}{cccccccc} 
 \hline
 Particle species & $B^0$ & $B^+$& $B^0_s$ & $\Lambda_b$ & $B_c^+$ & $c \overline{c}$ & $\tau^-\tau^+$ \\
   \hline

      Yield ($\times 10^9$)    & $310$ &  $310$ &  $75$  & $65$ & $1.5$ & $600$ & $170$ \\
  \hline
\end{tabular} 
\label{tab:flavouryields}
\end{table}

\section{Executive summary}

The study of the rare decays of $b$-flavoured hadrons occupies nowadays the front scene but the continued effort to accurately test the KM paradigm, one of the pillars of the SM, will remain in the future a must do of the flavour physics experimental program and shall conserve the attention of the community. We touched in this survey article the vibrant experimental projects that will shape the field in the next decades. The Belle II and LHCb upgrades programs, and in a farer and desirable future, the experiments at a high-energy $e^+e^-$ circular collider,  shall bring the precision on $C\!P$-violating observables to unprecedented levels. The anticipated subdegree accuracy on the CKM angles makes one illustration. The completeness of the study the CKM profiles requires that similar progresses are achieved on the $C\!P$-conserving quantities that this article focused on. In that respect, to make full use of the future experimental facilities precision, theoretical progress is mandatory both in perturbative and non-perturbative corrections. Since observables are typically products of short distance coefficients and long distance matrix elements, uncertainties have to be reduced in both factors.
Such a progress in theory is very challenging but doable if there are enough resources provided.

\section*{Acknowledgment}
The authors would like to thank Matthew Kirk and Oliver Witzel for helpful discussions.

\setboolean{inbibliography}{true}
\addcontentsline{toc}{section}{References}
\bibliographystyle{LHCb}
\bibliography{references}
\setboolean{inbibliography}{false}

\end{document}